\RequirePackage{fix-cm}
\documentclass[smallextended]{svjour3}       
\smartqed  
\usepackage{graphicx}
\usepackage{epstopdf}
\journalname{Journal of Low Temperature Physics}
\begin{document}

\title{Exciton-polariton solitons in a semiconductor microwire of finite size
}


\author{E. Nji Nde Aboringong \and
        Isaiah N. Ndifon \and
        Alain M. Dikand\'e
}


\institute{E. Nji Nde Aboringong \at
              Laboratory of Research on Advanced Materials and Nonlinear Sciences (LaRAMaNS), Department of Physics, Faculty of Science, University of Buea, P.O. Box 63 Buea, Cameroon \\
              \email{nji.edison@ubuea.cm}           
           \and
           Isaiah N. Ndifon \at
              Laboratory of Research on Advanced Materials and Nonlinear Sciences (LaRAMaNS), Department of Physics, Faculty of Science, University of Buea, P.O. Box 63 Buea, Cameroon \\
              \email{ndifon.isaiah@ubuea.cm}
               \and
           Alain M. Dikand\'e \at
              Laboratory of Research on Advanced Materials and Nonlinear Sciences (LaRAMaNS), Department of Physics, Faculty of Science, University of Buea, P.O. Box 63 Buea, Cameroon\\
              \email{dikande.alain@ubuea.cm}
}

\date{Received: date / Accepted: date}

\maketitle

\begin{abstract}
Exciton-polariton {\bf solitons} are strongly nonlinear quasiparticles composed of coupled exciton-photon states
due to the interaction of light with matter. In semiconductor microcavity systems such as semiconductor micro and nanowires, polaritons are characterized by a negative mass which when combined with the repulsive nonlinear exciton-exciton interaction, leads to the generation of bright polariton solitons. In this work we investigate the dynamics of {\bf bright exciton-polariton solitons} in a finite-size microcavity waveguide, for which radiative losses are assumed balanced by the external pumping. An exact {\bf bright-soliton} solution to the model equations of motion, consisting of {\bf a} periodic train of polariton pulses, is obtained in terms of Jacobi elliptic functions. Exact analytical expressions corresponding to the energies of both photonic and excitonic components of {\bf the pulse train} are found. Results suggest that the size (i.e. the length) of a microwire waveguide plays a relevant role in obtaining a quantitative estimate of the energy that could be conveyed by polariton solitons propagating in the medium.
\end{abstract}

\section{Introduction}
\label{intro}
The interaction of light with matter has attracted a great deal of attention in theoretical and experimental studies in the past years. These studies find relevance in physical systems such as low-dimensional molecular crystals \cite{SI1,SI2,SI3,SI4,SI5,SI6,SI7,SI8,SI9,ex2,prim2001,prim2002,dak,daka}, where several interesting and related phenomena have been observed especially for finite-sized molecular lattices \cite{edison2,edison3}. From a general standpoint, the interaction of light with matter involves the dynamics of certain quasiparticles usually referred to as \emph{exciton-polaritons}. These quasiparticles are regarded as bound-states of excitons strongly coupled to photons, when matter interacts sufficiently strongly with light. Their hybrid light-matter nature gives them unique properties as opposed to their single constituent counterparts. For instance, due to the presence of photons in polaritonic quasiparticles, the latter have an effective mass which is approximately $10^{-5}$ times lighter than that of a bare electron \cite{poldev}. This low effective mass, coupled with their bosonic character, has motivated the use of polaritons to achieve nonlinear phenomena such as Bose-Einstein condensation (BEC) \cite{bec1,bec2}, superfluidity \cite{sup1,sup2} and polariton lasing \cite{pl1,pl2,pl3}.
\par
Exciton polaritons have equally been investigated in semiconductor microcavities \cite{mcw1,bdmcw2,mcw3}, these optical resonators are useful for confining light to a small volume, either by means of total internal reflection on the wire surface or by the use of highly reflective multilayered Bragg mirrors. The latter case corresponds to optical resonators like the Fabry-Perot resonator, which
when combined with a cavity with active medium such as (In,Ga)As quantum wells, forms the
basis of a wide range of devices like lasers, enhanced photodiodes, and light emitting diodes \cite{bdmcw2,dev1,dev2}. Such semiconductor microcavities can be grown with very high quality advanced growth techniques like molecular beam epitaxy (MBE) \cite{gibbs2011} or metal-organic chemical vapour deposition (MOCVD) \cite{bdmcw2}. Polaritons in semiconductor micro and nano wires \cite{nan1} have recently been at the heart of research, since their low dimensionality eases manipulation and guidability in integrated optical circuits \cite{guid}. These low-dimensional structures can be envisaged as semiconductor microcavities etched into micro- and nano- wire shaped structures.
\par
A characteristic feature of polaritons in semiconductor microcavity waveguides is the remarkable dispersion spectrum of these quasiparticles \cite{newa,newb,newc,newd}. In the strong coupling regime, the dispersion curve consists of two branches (fig.1): an Upper Polariton Branch (UPB) which is dominantly photonic and a Lower Polariton Branch (LPB) which is almost completely excitonic at high values of the in-plane wave vector. A salient property of the LPB curve is that it contains points of inflexion, where the polariton effective mass (second order dispersion) changes sign \cite{micro1}. It has been established theoretically \cite{e2009} and observed experimentally \cite{sich2011} that the negative effective mass of polaritons coupled with repulsive polariton-polariton interactions leads to the formation of bright polariton solitons.
\par
In a recent study, the propagation of bright polariton pulses in a pump-free microcavity waveguide has been investigated \cite{e2019}. Analytical solutions for moving polariton solitons were proposed and related effects such as soliton compression, fission and emission of the backward Cherenkov radiation were highlighted. In contrast to Schr\"{o}dinger solitons, polariton pulses of different energy were found to separate during propagation.

However, from a theoretical perspective the bright polariton soliton solutions proposed in most recent works are localized pulses with a vanishing shape at the boundaries of a microwire cavity with an infinite length. This becomes a short coming in the context of physical systems with finite sizes, where the length of the propagation medium is expected to yield an important contribution to a quantitative estimate of the energy conveyed by these pulses.
\par
In this work, we revisit the theoretical analysis of the propagation of bright polariton pulses in microcavity wire waveguides \cite{e2019}. As a matter of fact, because microcavity wire waveguides are typically small-sized materials, they are characterized by well defined lengths that should be explicitely taken into consideration when investigating the dynamics of elementary excitations such as exciton-polariton quasiparticles. In this study we find solutions to the mathematical model using boundary conditions consistent with the finite size of the material, which lead to periodic structures representing trains of bright (i.e. pulse) polariton solitons. In section \ref{sec2} we introduce the mathematical model and examine the dispersion properties of polariton. In section \ref{sec3}, analytical solutions to the equations of motion are obtained in terms of periodic trains of bright polariton pulses with identical amplitudes and average widths. In section \ref{sec4}, exact analytical expressions corresponding to the energies conveyed by both the photonic and excitonic components of the bright pulse train are obtained and analysed. The impact of the finite length of the propagation medium, on the phase difference of the cavity field and the exciton-polariton field is examined in section \ref{sec5}. The work ends with a conclusion.

\section{Equations of motion and linear polariton dispersion}
\label{sec2}
We consider a one-dimensional micro (or nano) cavity wire {\bf of a finite size $L$}, in which the strong coupling of excitons and micro cavity photons gives rise to polaritons. Here, photons are generated coherently by an external pump beam with amplitude $E_p$ and transverse momentum $k_p$. Such an exciton-polariton system can be described by the following coupled set of equations \cite{bdmcw2,micro1,e2019}:
\begin{equation}\label{1}
  \frac{{\partial E}}{{\partial t}} - iD\frac{{\partial^2} E}{{\partial} x^2} +(\gamma_{ph} - i\Omega_{ph})E = i{\Omega_R}\Psi + {E_p}e^{i{k_p}x},
\end{equation}
\begin{equation}\label{2}
  \frac{\partial \Psi}{{\partial t}} - id\frac{{\partial^2} \Psi}{\partial x^2} +(\gamma_{ex} - i\Omega_{ex} + ig|\Psi|^2)\Psi = i{\Omega_R}E,
\end{equation}
where $E$ is the cavity field and $\Psi$ is the excitonic polarization. $\Omega_R$ is the Rabi frequency and $g$ is the exciton-exciton interaction coefficient. The terms $\Omega_{ph,ex}$ represent the detuning of the pump frequency ($w_p$) from the cavity ($w_{ph}$) and exciton ($w_{ex}$) resonances. The  terms $\gamma_{ph}$ and $\gamma_{ex}$ represent the cavity and exciton damping constants respectively. The coefficients $D$ and $d$ are defined in terms of the effective masses ($m_{ph,ex}$) of the cavity photons and excitons respectively as $D = {\hbar}/{2m_{ph}}$ and $d = {\hbar}/{2m_{ex}}$, where $m_{ex}\gg m_{ph}$.
\par
Disregarding nonlinear terms as well as losses and defining the cavity field and excitonic polarization, respectively, as:
\begin{eqnarray}
   E(x)= a(k)e^{i(kx - \Omega(k)t)}, \nonumber \\
  \Psi(x)= b(k)e^{i(kx - \Omega(k)t)},
\end{eqnarray}
where $k$ is the momentum along the waveguide axis, the linear polariton dispersion relation is given as:
\begin{eqnarray}\label{disprel}
  \Omega_{\pm}(k) &=& \frac{(D+d)k^2 - (\Omega_{ph} + \Omega_{ex})}{2} \nonumber \\
   &\pm&  \sqrt{{\Omega_{R}}^2 + \frac{1}{4}[(D - d)k^2 - (\Omega_{ph} - \Omega_{ex})]^2}.
\end{eqnarray}
\begin{figure}[h!]
\centering
\includegraphics[width=4.in]{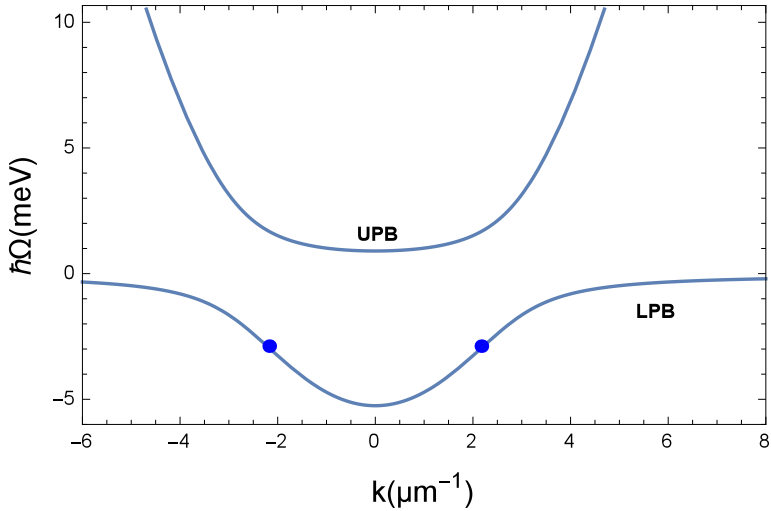}
  \caption{(Color online) Polariton dispersion relation showing the upper polariton branch (UPB) and the lower polariton branch (LPB). The two dots on the LPB are the inflection points. Modelling parameters are taken from ref.\cite{e2019}; $\hbar \Omega_R = 2.25meV$, $\hbar \Omega_{ph} = 4.2meV$, $\hbar \Omega_{ex} = 0$,
  $\hbar D = 0.65meV \mu m^2$, $\hbar d = 0.1\mu eV \mu m^2$. For the above modelling parameters, inflexion points occur at $k$-values $\pm 2.17\mu m^{-1}$.
 }
  \label{fig:Fig1}
\end{figure}
As illustrated in fig.\ref{fig:Fig1}, the polariton dispersion curve is characterized by an upper ($\Omega_{+}$) and a lower branch ($\Omega_{-}$). The upper branch is largely photonic while the lower branch is almost completely excitonic at high values of the momentum wave vector. In the strong coupling regime, the LPB contains sections of both positive and negative curvature (inflexion points) along one spatial direction. At these points, the polariton effective mass transits from positive to negative values at some critical values of the momentum wave vector. In fig.\ref{fig:Fig1}, the inflexion points are indicated by the two dots along the LPB. The negative polariton effective mass coupled with repulsive exciton-exciton interaction results in the formation of nonlinear localized polariton pulses.
\section{Soliton solutions}
\label{sec3}
In a bid to obtain exact soliton solutions to the equations of motion, we follow the approach used in ref. \cite{e2019}. However we shall focus our attention on the generation of propagating periodic polariton soliton trains, instead of localized solitary-wave structures. The former are more appropriate in the present context, given that they are consistent with periodic boundary conditions imposed by {\bf finite values of the size $L$ of microwire waveguides}. We begin by linearizing the parabolic dispersion relation of the photons and excitons around the transverse momentum $k_p$ of the pump beam, such that:
\begin{eqnarray}\label{5}
   E(x, t)= A'(x, t)e^{ik{_p}x}, \nonumber \\
  \Psi(x, t)= B'(x, t)e^{ik{_p}x}.
\end{eqnarray}
We seek polariton soliton solutions whose existence in semiconductor microcavities stems from the balance between losses and external pumping. To this end, we shall simplify our model by assuming that energy is conserved such that the system can be treated as `sourceless' $(E_p = 0)$ and `lossless' $(\gamma_{ph} = \gamma_{ex} = 0)$. Although in reality true physical systems must be characterized by sources and losses, we can interpret our approximate model as a state of the system after pumping is removed and significant losses have not presented themselves. Substituting equations (\ref{5}) into equations (\ref{1}) and (\ref{2}) and disregarding second-order derivative terms (where we have assumed pulses with sufficiently narrow spectra), we obtain:
\begin{equation}\label{6}
 i\frac{\partial A'}{\partial t} + 2iDk_{p}\frac{\partial A'}{\partial x} + (\Omega_{ph} - {Dk_{p}}^2)A' = -\Omega_RB',
\end{equation}
\begin{equation}\label{7}
 i\frac{\partial B'}{\partial t} + (\Omega_{ex} - g|B'|^2)B' = -\Omega_RA'.
\end{equation}
It is important to note that the last equation has been introduced to reflect the fact that before the pump beam was removed, it supplied purely photonic states.
By carrying out the following frequency and momentum transformations on the excitonic and photonic fields:
\begin{eqnarray}\label{trans}
   A'(x,t) = Ae^{i(k_s x+(\Omega_{ph} - D{k_{p}}^2 - \delta_s)t)}, \nonumber \\
  B'(x,t) = Be^{i(k_s x+(\Omega_{ph} - D{k_{p}}^2 - \delta_s)t)},
\end{eqnarray}
the system of equations in (\ref{6}) and (\ref{7}) can be cast into the form:
\begin{eqnarray}
  i\left( \frac{\partial A}{\partial t} + \eta \frac{\partial A}{\partial x}  \right) + \Omega_R B &=& 0, \label{9} \\
  i\frac{\partial B}{\partial t} + \Delta B - g|B|^2 B + \Omega_R A &=& 0.\label{10}
\end{eqnarray}
The quantity $\eta = 2k_p D$ is the group velocity of the cavity mode and $k_s$ is the free momentum parameter. We define the soliton frequency, $\Delta$, for a given value of the transverse momentum $k_p$ as $\Delta =\Omega_{ex} - \Omega_{ph} + D{k_{p}}^2 + \delta_s$. The term $\delta_s$ represents some constant soliton frequency shift. We introduce the new coordinate $z = t - x/v_s$, into equations (\ref{9}) and (\ref{10}), where $v_s$ is the velocity of propagation of the soliton. This results in the following system of equations characterizing the polariton solitary-wave profile:
\begin{equation}\label{11}
  i\sigma\frac{\partial A}{\partial z} - \Omega_R B = 0,
\end{equation}
\begin{equation}\label{12}
  i\frac{\partial B}{\partial z} + \Delta B - g|B|^2 B + \Omega_R A = 0,
\end{equation}
where $\sigma = \eta/v_s - 1$.
The excitonic and photonic fields represented by the last coupled set of equations above possess integrals of motion which can be expressed as \cite{peschel97}:
\begin{eqnarray}
  C_1 &=& \int dz |A|^2, \label{IOM1} \\
  C_2 &=& \frac{i}{2} \int dz \left[B^{*}\frac{\partial B}{\partial z} - B\frac{\partial B^{*}}{\partial z} \right], \label{IOM2}
\end{eqnarray}
where $C_1$ and $C_2$ represent the energy and quasi Hamiltonian respectively. Consequently, our system obeys two conservation laws by which the ordinary differential equations (\ref{11}), (\ref{12}) become integrable. These are given by:
\begin{eqnarray}
 -\sigma |A|^2 +|B|^2 &=& Q, \label{Q}\\
   \frac{\Omega_R}{2}(B^*A + BA^*) +\frac{\Delta}{2}|B|^2 - \frac{g}{4}|B|^4 &=& H, \label{H}
\end{eqnarray}
where $Q$ and $H$ are integration constants.
We define:
\begin{eqnarray}
  A &=& \rho_{A} \exp({i\varphi_A}), \nonumber \\
  B &=& \rho_{B} \exp({i\varphi_B}), \label{AB}
\end{eqnarray}
where $\rho_{A,B}$ and $\varphi_{A,B}$ are real amplitudes and phases of the complex fields $A$ and $B$ respectively. Introducing the intensity of the pulse $I = \rho^2_A$, and making use of the conservation laws (\ref{Q}) and (\ref{H}), we obtain from the field equations a first integral equation given as:
\begin{equation}\label{FI}
  \left(\frac{dI }{dz} \right)^2 = J I^4 + G I^3 + F I^2 + M I + N.
\end{equation}
The coefficients $J$, $G$, $F$, $M$ and $N$ are defined by the following relations:
\begin{eqnarray}
  J &=& -\frac{g^2 \sigma^2}{4}, \nonumber \\
  G &=& g\Delta \sigma - g^2Q\sigma, \nonumber \\
  F &=& \frac{1}{2\sigma}[8\Omega^2_R - 2\sigma \Delta^2 +6g\Delta Q\sigma - 3g^2Q^2\sigma - 4gH\sigma], \nonumber \\
  M &=& \frac{1}{\sigma^2}[4\Omega^2_R Q - 2Q\Delta^2\sigma + 3g\Delta Q^2\sigma + 4\Delta H\sigma \nonumber \\
  &-& g^2Q^3\sigma - 4gHQ\sigma], \nonumber \\
  N &=& \frac{1}{4\sigma^2}[4g\Delta Q^3 + 16\Delta QH - 4\Delta^2Q^2 - g^2Q^4 \nonumber \\
  &-& 8gHQ^2 - 16H^2].
\end{eqnarray}
Solitary-wave solutions to the field equations are characterized by vanishing field intensities
{\bf as $z\rightarrow \pm \infty$, thus requiring zero values of the integration constants $Q$ and $H$}. In this latter case, a strict proportionality exists between excitonic and photonic components of the polariton soliton as reflected in the quasi energy conservation relation, which is transformed to:
\begin{equation}\label{qer}
 {\rho^2_B} = \sigma {\rho^2_A}.
\end{equation}
It has been shown \cite{e2019} that for zero values of the integration constants $Q$ and $H$, equation (\ref{FI}) admits single pulse propagating soliton solutions of the form:
\begin{equation}\label{SSS}
  \rho^2_A = \frac{1}{g\Omega_R\sigma^{\frac{3}{2}}} \frac{4\Omega^2_R - \sigma \Delta^2}{\cosh \left[\sqrt{\frac{4\Omega^2_R - \sigma \Delta^2}{\sigma}(z-z_0)} \right] - \frac{\sqrt{\sigma}\Delta}{2\Omega_R}}.
\end{equation}
 The first integral equation  can be integrated in terms of elliptic functions, provided the condition $QH<0$ is imposed on the constants of integration $Q$ and $H$ in equation (\ref{FI}). This gives rise to finite non-zero values of the integration constants which are:
 \begin{eqnarray}
   Q^2 &=& \frac{\Delta^2}{g^2}, \nonumber \\
   H^2 &=& \frac{\Omega^4_R}{4g^2\sigma^2}. \label{finiteH}
 \end{eqnarray}
 For this last choice of integration constants, the first integral equation simplifies to:
 \begin{equation}\label{FIeq}
  \left(\frac{dI }{dz} \right)^2 = -J_0 I^4 + F_0 I^2 + N_0,
\end{equation}
where:
\begin{eqnarray}
  J_0 &=& \frac{g^2\sigma^2}{4}, \nonumber \\
  F_0 &=& \frac{5\Omega^2_R}{\sigma} + \frac{\Delta^2}{2}, \nonumber \\
  N_0 &=& -\frac{\Delta^4}{4g^2\sigma^2} + \frac{\Delta^2\Omega^2_R}{g^2\sigma^3}-\frac{\Omega^4_R}{g^2\sigma^4}.
\end{eqnarray}
Equation (\ref{FIeq}) admits a periodic bright soliton solution given by \cite{edison1,dik2010,dikbitha2018,fandio2015,fandiodik2017,mbieda2017}:
\begin{equation}\label{PSS}
  \rho^2_A = u_\kappa dn \left[\frac{x-v_st}{\lambda_\kappa},\kappa\right],
\end{equation}
which represents a pulse train of propagating polariton solitons, with identical pulse amplitudes $(u_\kappa)$ and average widths $(\lambda_\kappa)$ defined by the relations:
\begin{eqnarray}
  u_\kappa &=& \sqrt{\frac{F_0}{(2-\kappa^2)J_0}}, \\
  \lambda_\kappa &=& \sqrt{\frac{(2-\kappa^2)v^2_s}{F_0}}.
\end{eqnarray}
The function dn() is a Jacobi elliptic function of modulus $\kappa$ \cite{steg}, lying within the interval $0<\kappa\leq1$. As $\kappa \rightarrow 0$, the dn() function tends to a sine function, while as $\kappa \rightarrow 1$, the dn() function tends to a sech() function. \par 
{\bf It is instructive to stress that Jacobi elliptic functions are periodic in their arguments. In physical contexts where they interfer occuring as soliton solutions to nonlinear wave equations, their periodic features make them ideal wave patterns forming when the propagation media are either of finite size or display some spatial periodicity. In these contexts it is well established \cite{dik2010,dikbitha2018,dikd,dika,dikkof95} that the period of the Jacobi elliptic soliton solutions is equivalent to the size of the propagation medium $L$. Thus, for the specific pulse-train solution obtained in (\ref{PSS}) the period will be}:
\begin{equation}\label{period}
 L = 2K(\kappa)\lambda_\kappa,
\end{equation}
where $K(\kappa)$ is the elliptic integral of first kind. In the limit $\kappa\rightarrow1, L\rightarrow\infty$, the pulse train solution is transformed into a propagating solitary wave whose profile is calculated as:
\begin{equation}\label{K1S}
   \rho^2_A = \sqrt{\frac{F_0}{J_0}} sech[\sqrt{\frac{F_0}{v^2_s}}(x-v_st)].
\end{equation}
Alternatively, the single soliton solution (\ref{K1S}) can be obtained directly by integrating the simplified first integral equation (\ref{FIeq}) using the boundary conditions $I(z\rightarrow\pm\infty) \rightarrow 0$.
This last limit is descriptive of a microwire waveguide of infinite length in which bright polariton pulses with shape profiles of the form (\ref{K1S}) propagate. Therefore, the periodic solution (\ref{PSS}) can be regarded as a nonlinear finite-length solution \cite{dikkof95,dik96,dik1999} of (\ref{FI}), whose period coincides with the length of medium. In this regard the quantity $L$ corresponds to the length of the propagation medium. From formula (\ref{period}), the length $L$ of the propagation medium is proportional to the average widths of the individual pulses composing the periodic pulse train (\ref{PSS}).
\begin{figure}[h!]
\centering
\includegraphics[width=4.5in]{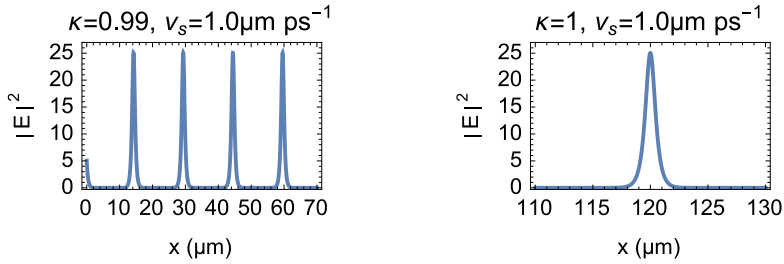}
  \caption{(Color online) Snapshots of the polariton pulse train (\ref{PSS}) propagating with velocity $v_s = 1.0\mu m$ $ps^{-1} $  for two values of $\kappa$: $\kappa = 0.99$ on the left graph and $\kappa = 1$ on the right graph. Other parameter values are; $\hbar \Omega_R = 2.25meV$, $\hbar \Omega_{ph} = 3.7meV$, $\hbar \Omega_{ex} = -0.5meV$, $\hbar D = 0.65meV \mu m^2$, $\hbar d = 0.1\mu eV \mu m^2$, $\hbar g = 0.05meV \mu m$, $k_p = 3.98\mu m^{-1}$.
 }
  \label{fig:Fig2}
\end{figure}
To further illustrate the qualitative behaviour of the periodic soliton solution obtained in formula (\ref{PSS}), we sketch in figure (\ref{fig:Fig2}) the polariton pulse train propagating at a velocity $v_s = 1.0\mu m$ $ps^{-1} $, for $\kappa = 0.99$ (left graph) and $\kappa = 1$ (right graph). As expected, the single pulse solitary-wave profile is recovered from the pulse train in the limit $\kappa\rightarrow1$ $(L \rightarrow\pm \infty)$. It should be stressed that the single pulse solution obtained in formula (\ref{K1S}) is different from that obtained in (\ref{SSS}) because in each of the cases, the integration constants $Q$ and $H$ were chosen differently. Indeed in the previous study \cite{e2019}, $Q$ and $H$ were chosen to be zero in the context of a microwire of infinite length. In our present context, we have chosen non-zero values of the integration constants as defined by formula (\ref{finiteH}) which enable us obtain the elliptic first integral equation of fourth degree (\ref{FIeq}).
\begin{figure}[h!]
\centering
\includegraphics[width=4.1in]{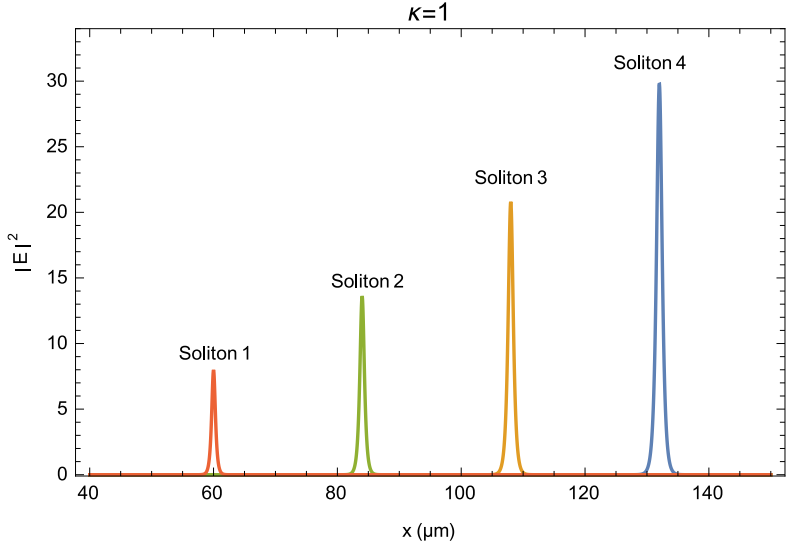}
  \caption{(Color online) Snapshots of the polariton pulse train (\ref{PSS}) with $\kappa = 1$ for four different values of the velocity of propagation $v_s = 0.4\mu m$ $ps^{-1}$, $0.7\mu m$ $ps^{-1}$, $0.9\mu m$ $ps^{-1}$, $1.1\mu m$ $ps^{-1}$, for solitons $1 \rightarrow 4$ respectively. Other parameter values are; $\hbar \Omega_R = 2.25meV$, $\hbar \Omega_{ph} = 3.7meV$, $\hbar \Omega_{ex} = -0.5meV$, $\hbar D = 0.65meV \mu m^2$, $\hbar d = 0.1\mu eV \mu m^2$, $\hbar g = 0.05meV \mu m$, $k_p = 3.98\mu m^{-1}$.
 }
  \label{fig:Fig3}
\end{figure}

\begin{figure}[h!]
\centering
\includegraphics[width=4.6in]{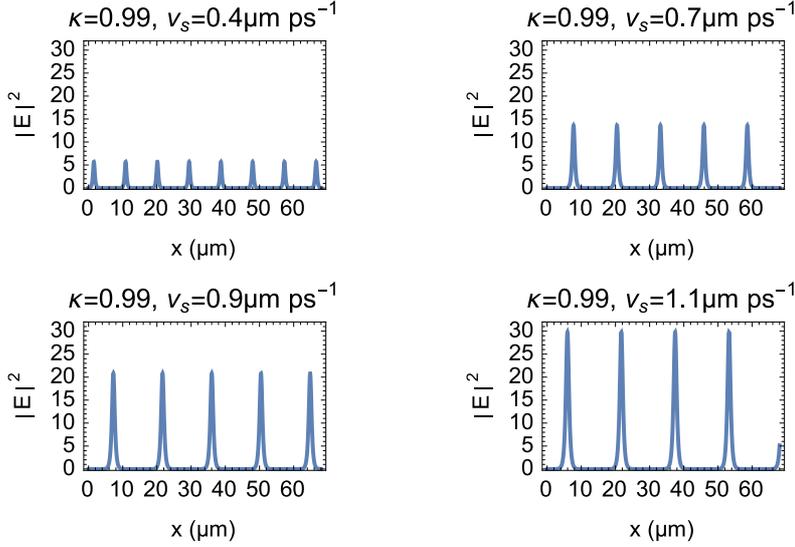}
  \caption{(Color online) Snapshots of the polariton pulse train (\ref{PSS}) with $\kappa = 0.99$ for four different values of the velocity of propagation $v_s = 0.4\mu m$ $ps^{-1}$, $0.7\mu m$ $ps^{-1}$, $0.9\mu m$ $ps^{-1}$, $1.1\mu m$ $ps^{-1}$. Other parameter values are; $\hbar \Omega_R = 2.25meV$, $\hbar \Omega_{ph} = 3.7meV$, $\hbar \Omega_{ex} = -0.5meV$, $\hbar D = 0.65meV \mu m^2$, $\hbar d = 0.1\mu eV \mu m^2$, $\hbar g = 0.05meV \mu m$, $k_p = 3.98\mu m^{-1}$.
 }
  \label{fig:Fig4}
\end{figure}
By definition, a soliton is a self-reinforcing wave packet propagating at a constant velocity and characterized by a robust shape profile. In figures (\ref{fig:Fig3}) and (\ref{fig:Fig4}), we sketch the shape profiles of four different solitons of the form (\ref{PSS}) propagating at different  velocities, for $\kappa = 1$ (figure \ref{fig:Fig3}) and $\kappa = 0.99$ (figure \ref{fig:Fig4}). Soliton $1$ represents a slowly moving soliton while soliton $4$ is a fast moving soliton. As the velocity of propagation is gradually increased from solitons $1\rightarrow4$, we observe  a huge increase in the amplitudes of the corresponding solitons (figure \ref{fig:Fig3}). Similarly, from figure (\ref{fig:Fig4}), the amplitudes of the identical pulses in the soliton train are greatly increased as the velocity is increased. This huge increase in soliton amplitudes is telltale of the corresponding increase in the energy conveyed by the soliton. Thus the faster the soliton, the greater the energy it conveys.
\section{Soliton energy in microwires of finite length}
\label{sec4}
For a microwire waveguide of finite-length $L$ as defined by the relation (\ref{period}), the total energy, $W$, conveyed by the polariton pulse train is given by the exact expression:
\begin{equation}\label{totalenergy}
  W = (1 + \sigma)W_A + \frac{2\Delta}{g} K(\kappa)\lambda_\kappa.
\end{equation}
The term $W_A$ represents the energy conveyed by the photonic component of the soliton train (\ref{PSS}) and is defined as:
\begin{eqnarray}
  W_A &=& \int_{-\frac{L}{2}}^{\frac{L}{2}} \rho^2_A dz, \nonumber \\
   &=& \frac{2\pi v_s}{g\sigma}.
\end{eqnarray}

On the other hand, the energy, $W_B$, conveyed by the excitonic component of the soliton train can be expressed in terms of $W_A$ using the relation:
\begin{equation}\label{WB}
  W_B = \sigma W_A + \frac{2\Delta}{g} K(\kappa)\lambda_\kappa.
\end{equation}
\begin{figure}[h!]
\centering
\includegraphics[width=3.1in]{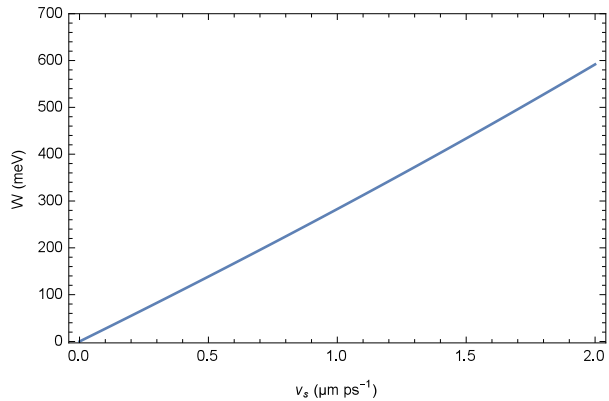}
  \caption{(Color online) Soliton energy $W$, versus the velocity of propagation $v_s$. Other parameter values are; $\hbar \Omega_R = 2.25meV$, $\hbar \Omega_{ph} = 3.7meV$, $\hbar \Omega_{ex} = -0.5meV$, $\hbar D = 0.65meV \mu m^2$, $\hbar d = 0.1\mu eV \mu m^2$, $\hbar g = 0.05meV \mu m$, $k_p = 9.5\mu m^{-1}$, $\kappa = 0.75$.
 }
  \label{fig:Fig6}
\end{figure}
From formula (\ref{totalenergy}), the total energy $W$, conveyed by the polariton pulse train is closely linked to the velocity of the pulse train and inversely proportional to the strength of the nonlinear exciton-exciton interaction. This dependence is illustrated in figure (\ref{fig:Fig6}), where we plot the soliton energy, $W$, versus the velocity of propagation, $v_s$. We observe a steady increase in the energy conveyed by the soliton train as the velocity of propagation is increased. This confirms the fact that fast moving polariton solitons convey a greater amount of energy and require finite-sized microwire waveguides of greater lengths to support their existence.

\section{The phase difference}
\label{sec5}
The phase difference $\varphi = \varphi_A - \varphi_B$ between the complex fields can be evaluated by making use of the relations (\ref{AB}) in the conservation laws (\ref{Q},\ref{H}). For the same choice of non-zero finite values of the integration constants, $Q$ and $H$, this leads to the expression:
\begin{eqnarray}\label{phaserel}
  \Omega_R\rho_A\cos(\varphi_A &-& \varphi_B) = -\frac{\Omega^2_R}{2g\sigma}\left(\frac{\Delta}{g}+\sigma\rho^2_A\right)^{-\frac{1}{2}} \nonumber \\
  &-&\frac{\Delta}{2}\left(\frac{\Delta}{g}+\sigma\rho^2_A\right)^{\frac{1}{2}} + \frac{g}{4}\left(\frac{\Delta}{g}+\sigma\rho^2_A\right)^{\frac{3}{2}}. \nonumber \\
\end{eqnarray}
The individual phases $\varphi_{A,B}$, can be obtained by integrating
the real parts of the equations for the fields with the help of the last expression (\ref{phaserel}). Following this approach, the phase difference $\varphi = \varphi_A - \varphi_B$ can be expressed as follows:
\begin{eqnarray}
  \varphi(z) &=& \frac{\Delta z}{4} + \frac{\Delta^2 +2\Omega^2_R}{4g\sigma^2}\varphi_1(z)
   + \varphi_2(z) + \frac{\Omega^2_R}{2\sigma}\varphi_3(z), \nonumber \\
\end{eqnarray}
where:
\begin{eqnarray}
 \varphi_1(z) &=& \frac{\lambda_\kappa}{\kappa'v_su_\kappa}\arccos\left[cd\left(\frac{v_sz}{\lambda_\kappa},\kappa\right)\right], \nonumber \\
  \varphi_2(z) &=& \arcsin\left[sn\left(\frac{v_sz}{\lambda_\kappa},\kappa\right)\right], \nonumber \\
  \varphi_3(z) &=& -\frac{2\arctan\left[\sqrt{1 - \beta_\kappa} sc\left(\frac{v_sz}{\lambda_\kappa},\kappa\right)\right]}{(\Delta^2 - g^2\sigma^2u^2_\kappa)\sqrt{1 - \beta_\kappa}} \nonumber \\
   &+& \frac{\Delta\lambda_\kappa\Pi(\beta_\kappa; \phi,\kappa)}{v_s(\Delta^2 - g^2\sigma^2u^2_\kappa)}.
\end{eqnarray}
The function $sc()$ is another jacobi elliptic function which represents the ratio of the $sn()$ and $cn()$ functions with complementary modulus $\kappa' = \sqrt{1-\kappa^2}$. $\Pi(\beta_\kappa; \phi,\kappa)$ is the elliptic integral of third kind with modulus $\kappa$, amplitude $\phi$ and characteristic $\beta_\kappa$, given as:
\begin{equation}\label{beta}
  \phi= am\left(\frac{v_sz}{\lambda_\kappa},\kappa\right), \qquad \beta_\kappa = \frac{\kappa^2g^2\sigma^2u^2_\kappa}{g^2\sigma^2u^2_\kappa - \Delta^2},
\end{equation}
where $am()$ is the Jacobi amplitude function \cite{steg}.

\section{Conclusion}
\label{sec6}
 We have investigated the propagation of bright polariton solitons in finite-sized one-dimensional semiconductor micro-and nano wires. Such microwire structures can be grown directly through chemical vapour deposition, or etched into semiconductor microcavities. The existence of polariton solitons in these microcavity waveguides is hinged on the balance between external pumping and losses. Therefore, we have reviewed an approximate mathematical model which was previously shown to admit exact single soliton solutions for a lossless-pump-free microcavity wire of infinite length.

 In this work, we have obtained elliptic type solitary-wave solutions to the model equations of motion which represent a periodic train of bright polariton solitons with identical amplitudes and average widths. We found that the length of the propagation medium is indeed proportional to the period of the soliton train solution, thus giving rise to interesting implications on the energy conveyed by these solitons. Exact analytical expressions for the total energy conveyed by the polariton pulse train solutions were obtained for a finite length microwire waveguide. It was established that fast moving polariton solitons convey a greater amount of energy and require finite-sized microwire waveguides of greater lengths to support their existence. These results suggest that the length dimensions of a microwire waveguide play a relevant role in determining the quantitative estimate of the energy that could be conveyed by polariton solitons propagating in the medium. An expression for the phase difference between the photonic and excitonic complex fields was also obtained in terms of elliptic functions. {\bf To end, remark that in this study we were mainly interested in lossless \cite{ex2} exciton-polariton soliton structures in semiconductor nanowires. In some physical contexts (see e.g. ref. \cite{eg}), radiative losses due to the exciton coupling to photons can be so sizable that the soliton amplitude gets damped upon propagation, leading to dissipative solitons.} 

\begin{acknowledgements}
Work partially supported by the Ministry of Higher Education of Cameroon (MINESUP), within the framework of the "Research Modernization Allowances".
\end{acknowledgements}

%
 \section*{Conflict of interest}
The authors declare that they have no conflict of interest.


\end{document}